\documentclass[aps,12pt,notitlepage,a4paper]{revtex4}

\usepackage{amsmath}
\usepackage{amsfonts}
\usepackage{amssymb}
\usepackage{graphicx}
\usepackage{fancyhdr}
\usepackage{amsthm} 	
\usepackage{mathrsfs}
\usepackage{wrapfig}
\usepackage{txfonts} 
\usepackage{setspace}
\usepackage{color}
\usepackage{curves}
\usepackage{rotating}
\usepackage{psboxit}
\usepackage{bm}% bold math
\usepackage{bbm}
\usepackage{subfigure}
\usepackage{float}

\newcommand{\ket}[1]{\left|{#1}\right>}
\newcommand{\bra}[1]{\left<{#1}\right|}
\newcommand{\braket}[2]{\left<{#1}\left|{#2}\right.\right>}
\newcommand{\im}{\mathrm{i}}

\newtheorem{definition}{Definition}
\newtheorem{theorem}{Theorem}

\begin{document}

\title{Some attempts at proving the non-existence of a full set of mutually unbiased bases in dimension 6}

\author{Thiang Guo Chuan}
\email{gct26@cam.ac.uk}
\affiliation{Centre for Quantum Technologies, National University of Singapore,\\ 3 Science Drive 2, 117543, Singapore}

\date{28 September 2010}
\maketitle

\section{Outline}
A set of $k$ mutually unbiased bases (MUB) in $\mathbb{C}^N$ is a set of orthonormal bases whose basis vectors obey the following relations:
\begin{equation}
	\left|\braket{e_i^{\alpha}}{e_j^{\beta}}\right|=\left\{\begin{array}{ll}\delta_{ij}\;&\text{if}\,\, \alpha=\beta,\\ \frac{1}{\sqrt{N}}\;&\text{if}\,\, \alpha\neq\beta, \end{array}\right.
\end{equation}
where $\alpha,\beta=1,\ldots,k$ and $i,j=1,\ldots,N$.

We will describe a few approaches to the notorious problem of proving the (non)-existence of four mutually unbiased bases in dimension 6. These will include the notions of Grassmannian distance, quadratic matrix programming, semidefinite relaxations to polynomial programming, as well as various tools from algebraic geometry.

\section{Grassmannian distance}
A unit ket $\ket{e}$ in $\mathbb{C}^6$ is identified with the density operator $\ket{e}\bra{e}$ in the real affine space of unit-trace hermitian operators acting on $\mathbb{C}^6$. Choosing the completely mixed state $\frac{1}{6}\mathbb{I}$ as the origin, this becomes a vector space of real dimension $35$, which can also be viewed as the space of traceless hermitian operators on $\mathbb{C}^6$. Thus $\ket{e}$ corresponds to the real vector ${\bf{e}}=\ket{e}\bra{e}-\frac{1}{6}\mathbb{I}$. The inner product between two vectors is defined via the traceless matrices that they are paired with, that is, if ${\bf{m}}_i\leftrightarrow M_i-\frac{1}{6}\mathbb{I}, i=1,2$, then ${\bf{m}}_1\cdot{\bf{m}}_2=\frac{1}{2}\text{tr}\left\{(M_1-\frac{1}{6}\mathbb{I})(M_2-\frac{1}{6}\mathbb{I})\right\}$. Therefore, we can view the set of unit ket vectors as an embedded subset of $\mathbb{R}^{35}$.

An orthonormal basis of ket vectors $\{\ket{e_i}\}_{i=1,\ldots,6}$ corresponds to the vectors $\{{\bf{e}}_i\}_{i=1,\ldots,6}$, which together span a $5$-dimensional subspace \cite{durt10}. Furthermore, the condition of mutual unbiasedness, $\left|\braket{e_i}{f_j}\right|^2=\frac{1}{6}$, becomes an orthogonality condition, ${\bf{e}}_i\cdot{\bf{f}}_j=0$, between the subspaces representing the bases $\{
\ket{e_i}\}, \{|f_j\rangle\}$. The Grassmannian of $5$-planes in $\mathbb{R}^{35}$ can be made into a metric space as follows. Let $\Pi_i$ denote the orthogonal projector onto a $5$-plane. Then the function $D(\Pi_1,\Pi_2)^2=\frac{1}{2}\text{tr}\left\{(\Pi_1-\Pi_2)^2\right\}$ is the desired distance function. Note that $D^2\in [0,5]$, with the maximal distance attained iff $\Pi_1$ and $\Pi_2$ are mutually orthogonal. We can extend this notion of distance to an average distance, defined as a function of four rank-5 projectors: $\bar{D}(\Pi_1,\Pi_2,\Pi_3,\Pi_4)^2=\frac{1}{6}\sum_{i<j}{D(\Pi_i,\Pi_j)^2}$. We also have $\bar{D}^2\in[0,5]$, with the maximal average distance attained iff each pair of projectors is mutually orthogonal. 

Since an orthonormal basis in $\mathbb{C}^6$ can be represented by some rank-5 projector on $\mathbb{R}^{35}$, we can study the existence of mutually unbiased bases by looking those projectors. The idea is to maximize the function $\bar{D}^2$ over quartets of rank-5 projectors representing bases in $\mathbb{C}^6$. A global maxima that is strictly smaller than 5 then suffices to prove the non-existence of four mutually unbiased bases in $\mathbb{C}^6$. There are, however, major problems with this approach.

Firstly, there is the troublesome constraint that the $\Pi_i$ are rank-5 projectors which come from orthonormal bases in $\mathbb{C}^6$. Presumably, one parameterizes the $\Pi_i$ by regarding them as elements of the vector space of symmetric $35\times 35$ matrices (or simply as matrices with an additional symmetry constraint). Then, the rank-5 projector property is imposed by the constraints $\Pi_i^2-\Pi_i=0$ and $\text{tr}\left\{\Pi_i\right\}=5$. As a first simplification, we ignore the requirement that the $\Pi_i$ correspond to orthonormal bases. The objective function being maximized is quadratic in the parameters specifying the matrices $\Pi_i$. Altogether, we have at least a quadratically-constrained quadratic program (QCQP), which in the general case is $NP$-hard. Occasionally, if the quadratic forms involved are definite, one can use Schur complements to turn the QCQP into a semidefinite program, whose global maxima can of course be found. This is, unfortunately, not the case here. We might ask for something less, such as an upper bound for the global maxima, rather than its actual value. However, the bounds obtained so far have been trivial. One can also look at the Lagrange dual of this (primal) optimization problem, the reason being that the dual gives upper bounds to the primal, and furthermore, the dual is always convex. Strangely enough, the dual problem gives minimally trivial upper bounds, which roughly speaking, indicates that the non-convexity of the primal is crucial and must be taken into account if we wish to draw non-trivial conclusions. And we have not even checked that the projectors refer to bases. We cannot even hope to relax this last constraint, because we would be left with the equivalent geometrical problem of packing $5$-planes orthogonally in $\mathbb{R}^{35}$, for which it is trivially true that a maximum of exactly seven such planes can be fitted.

Of course, one can start directly from the bases of ket vectors $\{\ket{e_i}\}$, and then build the corresponding rank-5 projectors. However, the average-distance function $\bar{D}^2$ becomes quartic, which as an optimization problem is even more difficult than a quadratic one.

\section{Quadratic matrix programming and Dattorro's convex iteration}
Some consolation can be derived from the fact that a number of techniques exist for handling QCQP. In particular, semidefinite relaxations of QCQP have been studied for some time. Of interest here is a variant of this idea, which is called \emph{quadratic matrix programming} \cite{beck07} (QMP). This refers to nonconvex quadratic optimization problems of the following form:
\begin{equation}
\begin{array}{ll}
	\displaystyle{\mathop{\mbox{minimize}}_{X\in\mathbb{R}^{n\times r}}} & \text{tr}\left\{X^TA_0X\right\}+2\text{tr}\left\{B_0^TX\right\}+c_0 \\
	\text{subject to} & \text{tr}\left\{X^TA_iX\right\}+2\text{tr}\left\{B_i^TX\right\}+c_i=0, i=1,\ldots,k,
\end{array}\label{QMP}
\end{equation}
where $A_i\in\mathbb{R}^{n\times n}, B_i\in \mathbb{R}^{n\times r}, c_i\in\mathbb{R}, i=0,\ldots, k$. Note that the objective and constraint functions are quadratic matrix functions of order $r$. We shall use the abbreviation $f_i(X)=\text{tr}\left\{X^TA_iX\right\}+2\text{tr}\left\{B_i^TX\right\}+c_i$, and call $f_i$ homogeneous if $B_i=0_{n\times r}, c_i=0$.

The construction of a semidefinite relaxation to the QMP \eqref{QMP} begins by the process of homogenization. For $f_i$ as given above, the corresponding homogenized quadratic matrix function $f_i^H$ is defined by
\begin{equation}
	f_i^H(Y;Z)\equiv\text{tr}\left\{Y^TA_iY\right\}+2\text{tr}\left\{Z^TB_i^TY\right\}+\frac{c_i}{r}\text{tr}\left\{Z^TZ\right\}, Y\in\mathbb{R}^{n\times r}, Z\in\mathbb{R}^{r\times r},
\end{equation}
which is homogeneous, and can be represented by the matrix
\begin{equation}
	M(f_i)=\begin{pmatrix}
		A_i & B_i \\
		B_i^T & \frac{c}{r}\mathbb{I}_r
	\end{pmatrix}\longleftrightarrow f_i(Y;Z)=\text{tr}\left\{(Y,Z)^TM(f_i)(Y,Z)\right\}.
\end{equation}

The homogenized version of \eqref{QMP} reads
\begin{equation}
	\begin{array}{ll}
		\text{minimize} & f_0^H(Y;Z) \\
		\text{subject to} & f_i^H(Y;Z)=0, i=1,\ldots,k, \\
		{} & \psi_{ij}(Y;Z)=2\delta_{ij}, 1\leq i\leq j\leq r,
	\end{array}\label{QMPhomo}
\end{equation}
where $\psi_{ij}(Y;Z)=\text{tr}\left\{Z^T(E_{ij}^r+E_{ji}^r)Z\right\}$ with $E_{ij}^r$ defined as the $r\times r$ matrix with zeroes everywhere except for the $(i,j)$-th entry, which is $1$. The importance of \eqref{QMPhomo} lies in the fact that it is solvable precisely when the original QMP \eqref{QMP} is solvable, and in that case, the optimal values of both problems are equal \cite{beck07}. 

With the homogenized problem at hand, we can form a semidefinite relaxation as follows. We write $(Y;Z)$ as the single matrix variable $W\in\mathbb{R}^{(n+r)\times r}$, so that \eqref{QMPhomo} turns into
\begin{equation}
	\begin{array}{ll}
		\text{minimize} & \text{tr}\left\{M(f_0)WW^T\right\} \\
		\text{subject to} & \text{tr}\left\{M(f_i)WW^T\right\}=0, i=1,\ldots,k, \\
		{} & \text{tr}\left\{N_{ij}WW^T\right\}=2\delta_{ij}, 1\leq i\leq j\leq r,
	\end{array}\label{QMPhomo2}
\end{equation}
where
\begin{equation}
	N_{ij}=\begin{pmatrix}
		0_{n\times n} & 0_{n\times r} \\
		0_{r\times n} & E_{ij}^r+E_{ji}^r
	\end{pmatrix},
\end{equation}
and the cyclic property of the trace has been been used. Now, observe that $U\equiv WW^T$ is a symmetric, positive semidefinite $(n+r)\times(n+r)$ matrix, so we have, equivalently, the optimization problem
\begin{equation}
	\begin{array}{ll}
		\text{minimize} & \text{tr}\left\{M(f_0)U\right\} \\
		\text{subject to} & \text{tr}\left\{M(f_i)U\right\}=0, i=1,\ldots,k, \\
		{} & \text{tr}\left\{N_{ij}U\right\}=2\delta_{ij}, 1\leq i\leq j\leq r,\\
		{} & U\geq 0\\
		{} & \text{rank}(U)\leq r.
	\end{array}\label{QMPSDPR}
\end{equation}
Notice that, apart from the rank constraint, \eqref{QMPSDPR} is a semidefinite program.

Let us try to relate QMP to the MUB existence problem. We have already seen, from the Grassmannian distance approach, that the basis vectors making up the candidate MUBs should be specified individually, because it is not clear how to ensure that a rank-5 projector in $\mathbb{R}^{35}$ corresponds to a basis in $\mathbb{C}^6$. If we choose to specify ket vectors, the MU conditions invariably become quartic. Therefore, we choose a compromise, which is to specify rank-1 density matrices. Note that two such density matrices $\rho_1=\ket{e}\bra{e}, \rho_2=\ket{f}\bra{f}$ are mutually unbiased iff their Hilbert-Schmidt inner product $\text{tr}\left\{\rho_1\rho_2\right\}=\braket{e}{f}\braket{f}{e}=\frac{1}{6}$.

In order to reduce the number of variables in the QMP that we are about to build, we can consider MU constellations \cite{brierley08} rather than full bases. In the density operator picture, these are sets of pure states $\left\{\rho_i^\alpha\right\}$ which obey \begin{equation}
	\text{tr}\left\{\rho_i^\alpha\rho_j^\beta\right\}=\left\{\begin{array}{ll}
	\delta_{ij} & \text{if}\: \alpha=\beta, \\
	\frac{1}{6} & \text{if}\: \alpha\neq\beta.
	\end{array}\right.\label{MUconditions}
\end{equation}
A necessary condition for a set of four MUBs $\left\{\rho_i^\alpha\right\}_{i=1,\ldots,6}^{\alpha=1,\ldots,4}$ to exist in $\mathbb{C}^6$ is that every subset of these pure states also obey the MU conditions \eqref{MUconditions}. A MU constellation is labelled by a set of numbers $\{a_1,a_2,\ldots,a_k\}_6$, which indicates that there are $k$ groups of pure states, and that the $i$-th group comprises $a_i$ pairwise orthogonal pure states. One of the smallest constellation that is not known to exist is $\{5,3,3,3\}_6$. By applying a global unitary transformation (or a change in basis), we may assume that the five pure states in the first group are given by the diagonal matrices $E_{ii}^6, i=1,\ldots,5$. This leaves us to specify three groups of three pure states $\left\{\rho_i^\alpha\right\}_{i,\alpha=1,2,3}$, which we accomplish by specifying their real and imaginary parts (so eighteen $6\times6$ real matrices have to be specified). For instance, $\rho_i^\alpha=C_i^\alpha+\text{i}D_i^\alpha$. The matrix $X$ appearing in \eqref{QMP} will then be the vertical concatenation of these 18 matrices. 

It remains to verify that the objective function as well as all the constraints that we need to impose are in fact quadratic matrix functions. We will try to set up a feasibility problem, so the objective function is just the zero function. The first constraint will be that $C_i^\alpha+\text{i}D_i^\alpha$ is hermitian, which is equivalent to saying that $C_i^\alpha$ is symmetric and $D_i^\alpha$ is antisymmetric. Since the symmetric and antisymmetric subspaces of $\mathbb{R}^{6\times 6}$ are mutually orthogonal, we just require that the components of $C_i^\alpha$ (resp. $D_i^\alpha$) in a basis of antisymmetric (resp. symmetric) matrices vanish. These conditions are of the form $\text{tr}\left\{C_i^\alpha B_\text{antisymm}\right\}=0$ and $\text{tr}\left\{D_i^\alpha B_\text{symm}\right\}=0$, which are quadratic (linear, in fact).

Next, we ensure that $\rho_i^\alpha=C_i^\alpha+\text{i}D_i^\alpha$ is a rank-1 projector by imposing $\text{tr}\left\{\rho_i^\alpha\right\}-1=0$ and $(\rho_i^\alpha)^2-\rho_i^\alpha=0$. It is straightforward to see that these conditions are quadratic in $C_i^\alpha$ and $D_i^\alpha$. Finally, the inner product between $\rho_i^\alpha$ and $\rho_j^\beta$ reads
\begin{equation}
\text{tr}\left\{\rho_i^\alpha\rho_j^\beta\right\}=\text{tr}\left\{(C_i^\alpha+\text{i}D_i^\alpha)(C_j^\beta+\text{i}D_j^\beta)\right\}=\text{tr}\left\{C_i^\alpha C_j^\beta-D_i^\alpha D_j^\beta\right\},
\end{equation}
which is again quadratic. Therefore, the MU conditions can be imposed via quadratic matrix functions. 

Following the prescription described above, one arrives at a SDP with a rank constraint ($\text{rank}(U)\leq 6$), which is \emph{equivalent} to the original feasibility problem. In order to obtain a certificate of infeasibilty, we can do the following. Convert one of the orthogonality conditions in the MU constraints, say $\text{tr}\left\{\rho_1^1\rho_2^1\right\}=0$, to an objective function $f_0$. We see that $f_0$ is non-negative, and attains zero precisely when orthogonality, along with all the other constraints are fulfilled. That is, $\text{min}(f_0)=0$ iff the $\{5,3,3,3\}_6$ constellation exists. Therefore, if we can show that the global minimum of $f_0$ is strictly positive, then we are done.

Ordinarily, one advantage of a SDP formulation is that global bounds can be found, by considering the dual SDP problem, for instance. Thus, the only obstacle remaining is the rank constraint. There has been some work done on SDPs with rank constraints, but those are local methods which provide low-rank solutions that are not necessarily globally optimal. On the flip side, if we can find a low-rank solution at which $f_0$ attains the value $0$ (or if we can simply find a low-rank solution to the feasibility problem), then we have found the elusive $\{5,3,3,3\}_6$ constellation. With this in mind, we look at a recently developed and rather controversial method of handling SDPs with rank constraints. 

Dattorro has suggested the so-called convex iteration procedure \cite{dattorro10} to find low-rank solutions to rank-constrained SDPs. Consider a semidefinite program with $\mathcal{C}$ as its convex feasible set. Then the rank-constrained semidefinite feasibility problem has the form
\begin{equation}
	\begin{array}{ll}
		\displaystyle{\mathop{\mbox{find}}_{G\in\mathbb{S}^N}} & G \\
		\text{subject to} & G\in\mathcal{C}\\
		{} & G\geq 0\\
		{} & \text{rank}(G)\leq n.
	\end{array}\label{SDPrankconstrained}
\end{equation}
The feasible set in this case is the intersection of $\mathcal{C}$ with a certain subset of the positive semidefinite cone boundary, namely the positive matrices of rank $n$ or less. This is clearly a non-convex set, and may even be empty. The convex iteration method looks for a feasible solution in this intersection by considering the following two coupled SDPs. 
\begin{equation}
	\text{(SDP1)\qquad}\begin{array}{ll}
		\displaystyle{\mathop{\mbox{minimize}}_{G\in\mathbb{S}^N}} & \text{tr}\left\{GW\right\} \\
		\text{subject to} & G\in\mathcal{C}\\
		{} & G\geq 0,
	\end{array}\label{SDP1}
\end{equation}
and
\begin{equation}
	\text{(SDP2)\qquad}\begin{array}{ll}
		\displaystyle{\mathop{\mbox{minimize}}_{W\in\mathbb{S}^N}} & \text{tr}\left\{GW\right\} \\
		\text{subject to} & 0\leq W\leq \mathbb{I}_N \\
		{} & \text{tr}\left\{W\right\}=N-n.
	\end{array}\label{SDP2}
\end{equation}
In \eqref{SDP1}, $W$ is an optimal solution to \eqref{SDP2}; likewise in \eqref{SDP2}, $G$ is an optimal solution to \eqref{SDP1}. The feasible set in \eqref{SDP2},
\begin{equation}
	F_{N-n}^N=\left\{W\in\mathbb{S}^N : 0\leq W \leq \mathbb{I}_N, \text{tr}\left\{W\right\}=N-n\right\},
\end{equation}
is called the $(N-n)$-Fantope. It is the convex hull of the set of all rank-$(N-n)$ projectors. In fact, the extreme points of the $(N-n)$-Fantope are precisely the rank-$(N-n)$ projectors.

One proceeds to solve \eqref{SDP1} and \eqref{SDP2} iteratively, until local convergence of $\text{tr}\left\{GW\right\}$ to some non-negative value $\tau$ is established. It is easy to see that the iterations will generate a non-increasing sequence of values for $\text{tr}\left\{GW\right\}$, because if at a certain stage the feasible pair $(G',W')$ gives $\text{tr}\left\{G'W'\right\}=\tau'$, then $\tau'$ must bound the subsequent minimization problems from above. The monotone convergence theorem then guarantees that the sequence of values of $\text{tr}\left\{GW\right\}$ converges to some real number $\tau\geq 0$. Let us see what we can conclude if we actually have $\text{tr}\left\{G^*W^*\right\}=\tau=0$ for some pair $(G^*,W^*)$. In this case, we would have found a $G^*\in\mathcal{C}$ whose range is orthogonal to that of $W^*$. But $W^*$ belongs to the $(N-n)$-Fantope, and has a rank of at least $(N-n)$. Therefore, $\text{rank}(G^*)\leq n$, \emph{id est} we have established feasibility of \eqref{SDPrankconstrained}.

Returning to our MU constellation existence problem, we see that the convex iteration procedure allows a systematic way to establish feasibility. We simply construct the rank-constrained SDP for the constellation in question, and then use convex iteration to search for a low-rank feasible point. An actual numerical implementation readily confirms the existence of the largest known constellations such as $\{5,3,3,2\}_6$ and $\{5,5,5\}_6$, the latter representing three MUBs in $\mathbb{C}^6$. However, convex iteration fails to find $(G^*,W^*)$ with  $\text{tr}\left\{G^*W^*\right\}=0$ for the $\{5,3,3,3\}_6$ constellation. In fact, it appears that the local convergence to $\tau$ depends on the initialization of the iterative procedure. The best that has been achieved is $\tau=0.0022$, which is inconclusive. The problem is that it is not known when $\text{tr}\left\{G^*W^*\right\}=0$ will actually be achieved via convex iteration for a feasible rank-constrained SDP. Therefore, our failure to find $\tau=0$ confirms nothing about the existence of the $\{5,3,3,3\}_6$ constellation. Indeed, convex iteration cannot provide an infeasibilty certificate. It only adds to the suspicion that a set of four MUBs in $\mathbb{C}^6$ does not exist.

\section{Lasserre's semidefinite relaxations}
Global optimization is extremely difficult for nonconvex problems. Nevertheless, Lasserre has developed a remarkable approach to polynomial problems, by defining a sequence of semidefinite programming relaxations of increasing size which provide ever better approximations to the original polynomial problem \cite{lasserre01,lasserre08,lasserre09}. There is even a publicly available MATLAB implementation GloptiPoly 3 \cite{henrion09}. This implementation solves Generalized Problems of Moments (GPM), of which the following is a special case:
\begin{equation}
	\begin{array}{ll}
		\displaystyle{\mathop{\mbox{minimize}}_{\text{d}\mu}} & \int_{\mathbb{K}}{p_0(x)\,\text{d}\mu(x)}\\
		\text{subject to} & \int_{\mathbb{K}}{h_j(x)\,\text{d}\mu(x)}\geq b_j,\;j=1,2,\ldots,
	\end{array}
\end{equation}
where $b_j$ are real numbers and the measure $\text{d}\mu$ in $\mathbb{R}^n$ is supported on the semialgebraic set $\mathbb{K}$ defined by the polynomials $p_i$,
\begin{equation}
	\mathbb{K}=\left\{x\in\mathbb{R}^n : p_i(x)\geq 0, i=1,2,\ldots\right\}.\label{semialgebraic}
\end{equation}
GloptiPoly 3 carries out the GPM optimization via the moments of the measure $\text{d}\mu$, defined as
\begin{equation}
	y_\alpha=\int_{\mathbb{K}}{x^\alpha\,\text{d}\mu(x)},\;\alpha\in\mathbb{N}^n,
\end{equation}
where $\alpha$ are multi-indices labelling the moments.

A general constrained polynomial problem has the form
\begin{equation}
	\displaystyle{\mathop{\mbox{minimize}}_{x\in\mathbb{K}}}\,p_0(x),
\end{equation}
where the set $\mathbb{K}$ is defined by the given polynomial constraints as in \eqref{semialgebraic}. Lasserre has shown \cite{lasserre01} that the above polynomial problem can be cast as the moment problem
\begin{equation}
	\begin{array}{ll}
		\displaystyle{\mathop{\mbox{minimize}}_{\text{d}\mu}} & \int_{\mathbb{K}}{p_0(x)\,\text{d}\mu(x)}\\
		\text{subject to} & \int_{\mathbb{K}}\,\text{d}\mu=1,
	\end{array}\label{moment}
\end{equation}
i.e., the decision variable $\text{d}\mu$ is a probability measure supported on the set $\mathbb{K}$. A hierarchy of semidefinite programming relaxations is then constructed, whose sequence of optimal values converges to the true global optimal value of \eqref{moment} \cite{lasserre08,lasserre09}.

One can view the equations defining a MU constellation as a system of polynomial equations. Following \cite{brierley10}, we consider the simplest constellation that is known not to exist, which is $\{1,1,1,1\}_2$. Four real variables suffice to parameterize a candidate set of four vectors:
\begin{equation}
	\begin{pmatrix}
		1\\0
	\end{pmatrix},
	\frac{1}{\sqrt{2}}
		\begin{pmatrix}
		1\\1
	\end{pmatrix},
	\frac{1}{\sqrt{2}}
		\begin{pmatrix}
		1\\x_1+\im x_2
	\end{pmatrix},
	\frac{1}{\sqrt{2}}
		\begin{pmatrix}
		1\\x_3+\im x_4
	\end{pmatrix}.
\end{equation}
Five polynomial equations then serve as constraints defining the MU constellation,
\begin{equation}
	\begin{array}{l}
		p_1(x)\equiv x_1^2+x_2^2-1=0,\\
		p_2(x)\equiv x_3^2+x_4^2-1=0,\\
		p_3(x)\equiv (1+x_1)^2+x_2^2-2=0,\\
		p_4(x)\equiv (1+x_3)^2+x_4^2-2=0,\\
		p_5(x)\equiv (1+x_1x_3+x_2x_4)^2+(x_1x_4-x_2x_3)^2-2=0.
	\end{array}\label{1111eqn}
\end{equation}
Brierley and Weigert \cite{brierley10} proposed to take $(p_1(x))^2$ as the objective function, and solve the optimization problem
\begin{equation}
	\begin{array}{ll}
		\text{minimize} & (p_1(x))^2\\
		\text{subject to} & p_i(x)=0,\,i=2,\ldots,5.
	\end{array}\label{1111constopt}
\end{equation}
Then, it follows that the global minimum of problem \eqref{1111constopt} is zero iff there is a solution to $p_i(x)=0, i=1,\ldots,5$, iff the constellation $\{1,1,1,1\}_2$ exists. Now, the optimal objective value attained at each relaxation in Lasserre's hierarchy is a global lower bound for the true minimum of the original problem. Therefore, a strictly positive optimal objective value at any level of relaxation serves as a certificate proving the non-existence of the constellation $\{1,1,1,1\}_2$. In practice, the size of the SDP relaxations increases very rapidly as we progress through the hierarchy. In \cite{brierley10}, three level of relaxations were calculated for the case of $\{1,1,1,1\}_2$. It was also claimed that the first relaxation, which yielded an optimal objective value of $1.4038\times 10^{-8}$, provided the desired certificate. However, it seems unclear whether this is truly a strictly positive value, or an artifact of numerical errors. The second relaxation gives a less ambiguous answer: $0.5359$ is certainly positive. The authors then proceeded to attempt the same construction for the $\{5,5,5,1\}_6$ case, which proved to be intractable. In fact, \emph{generating} the first SDP relaxation was already too difficult, let alone solving it.

The SDP relaxations for constellations like $\{5,5,5,1\}_6$ and $\{5,3,3,3\}_6$ are so large because of the large number of variables present in the polynomials defining the constellation, as well as the relatively high degree of the polynomials, which are quartic. We have tried an alternative parameterization of $\{5,3,3,3\}_6$ by specifying density operators rather than vectors in $\mathbb{C}^6$. For this, we first note that we can choose the density operators for the first group in the constellation to be $E_{ii}^6, i=1,\ldots,5$, corresponding the the choice of the computational basis for $\mathbb{C}^6$. The sixth density operator in this group is then automatically determined to be $E_{66}^6$. The remaining $3\times 3$ density operators will be parameterized as follows:
\begin{equation}
	\rho_i^\alpha=\frac{1}{6}\begin{pmatrix}
		1&\bar{z}_{\alpha,i,1}&\bar{z}_{\alpha,i,2}&\bar{z}_{\alpha,i,3}&\bar{z}_{\alpha,i,4}&\bar{z}_{\alpha,i,5}\\
		z_{\alpha,i,1}&1&\bar{z}_{\alpha,i,6}&\bar{z}_{\alpha,i,7}&\bar{z}_{\alpha,i,8}&\bar{z}_{\alpha,i,9}\\
		z_{\alpha,i,2}&z_{\alpha,i,6}&1&\bar{z}_{\alpha,i,10}&\bar{z}_{\alpha,i,11}&\bar{z}_{\alpha,i,12}\\
		z_{\alpha,i,3}&z_{\alpha,i,7}&z_{\alpha,i,10}&1&\bar{z}_{\alpha,i,13}&\bar{z}_{\alpha,i,14}&\\
		z_{\alpha,i,4}&z_{\alpha,i,8}&z_{\alpha,i,11}&z_{\alpha,i,13}&1&\bar{z}_{\alpha,i,15}\\
		z_{\alpha,i,5}&z_{\alpha,i,9}&z_{\alpha,i,12}&z_{\alpha,i,14}&z_{\alpha,i,15}&1
	\end{pmatrix},\;\alpha,i=1,\ldots,3.
\end{equation}
The diagonal entries are $\frac{1}{6}$ because of the MU requirement between the first basis and the remaining nine $\rho_i^\alpha$. Therefore, $15\times 9=135$ complex numbers, or $270$ real numbers are required. Next, we have to ensure that $\rho_i^\alpha$ is a rank-1 projector. For the rank-1 condition, it is necessary and sufficient to check that the columns of $\rho_i^\alpha$ are linearly dependent, or equivalently, that the determinant of every $2$nd order minor vanishes. Actually, we can do even better; we only need to impose that every neighbouring $2$nd order minor vanishes. Since $\rho_i^\alpha$ is hermitian by parameterization, we do not even have to check the minors that involve only the upper triangular entries, since these will be duplicated by their complex-conjugate counterparts in the lower triangular sector. Then, the projection property follows from the fact that $\text{tr}\left\{\rho_i^\alpha\right\}=1$ is the single non-vanishing eigenvalue of $\rho_i^\alpha$. Therefore, the rank-1 projection property is ensured by imposing $15$ quadratic equations for each $\rho_i^\alpha$. To enforce the MU constraints between different groups of density operators in the constellation, $\text{tr}\left\{\rho_i^\alpha\rho_j^\beta\right\}=\frac{1}{6},\,\alpha\neq\beta,\, i,j=1,2,3$, an additional $27$ quadratic constraints must be specified.

The only remaining constraints are the orthogonality MU constraints 
\begin{equation}
	\text{tr}\left\{\rho_i^\alpha\rho_j^\alpha\right\}=\delta_{ij},\: \alpha,i,j=1,2,3.
\end{equation}
Since we have already ensured that $\rho_i^\alpha$ is a projector and hence positive semidefinite, we observe that these orthogonality MU constraints are simultaneously fulfilled if and only if the \emph{single} combined quadratic constraint,
\begin{equation}
p_0\left(\{\rho_i^\alpha\}\right)\equiv\displaystyle{\mathop{\sum_{\alpha}}_{i<j}}{\text{tr}\left\{\rho_i^\alpha\rho_j^\alpha\right\}}=0,
\end{equation}
is fulfilled. Hence we can take $p_0$ to be the objective function to be minimized subject to the remaining constraints. This has the nice property that the global minimum of $f_0$ is zero iff the MU constellation $\{5,3,3,3\}_6$ exists. For this quadratic optimization problem that we have described, Gloptipoly 3 is able to generate the first SDP relaxation, which is already huge. It cannot be handled on an ordinary desktop PC. Although this is a slight improvement, it is likely that this first relaxation will yield inconclusive results, if the smaller-sized problems are to be a guide.

It is worth noting that reducing the problem from a quartic one to a quadratic one comes at the cost of introducing more variables and constraint equations. On a positive note, QCQP is itself an active field of research, so there is some hope that the large QCQP problem that we have described above can actually be handled.

\section{Algebraic geometry and Gr\"{o}bner bases}
It might very well be that carrying out a polynomial optimization to prove the non-existence of a certain MU constellation is an overkill. Perhaps it is really unnecessary to optimize or solve the polynomial equations; we might be satisfied with just knowing how the solutions ``look like", or how many of them there are. In essence, what we have is a set of $N$ multivariate polynomial equations $\{p_i(x)=0\}_{i=1\ldots,N}$ defining a MU constellation, and what we are interested in are the solutions, if any, to these equations. This leads us to the field of algebraic geometry, which is replete both with elegant theorems and frustrating open problems.

The polynomials $p_i$ belong the the ring of polynomials over $\mathbb{R}$ in variables $x_1,\ldots,x_n$, which is denoted by $\mathbb{R}[x_1,\ldots,x_n]$. Let $S$ be the set $\{p_i\}_{i=1,\ldots,N}$. Then the central object of interest is the set of common zeros of $S$ in $\mathbb{R}^n$, i.e. the (real) variety $V_\mathbb{R}(S)\equiv\left\{x\in\mathbb{R}^n : p(x)=0\: \forall p\in S\right\}$. In this definition, the MU constellation defined by $S$ exists if and only if $V_\mathbb{R}(S)\neq\emptyset$. Now, \emph{real} algebraic geometry is notoriously difficult, so we look for complex zeros instead. That is, we consider $S$ as a subset of $\mathbb{C}[x_1,\ldots,x_n]$ and study the complex variety $V_\mathbb{C}(S)\equiv\left\{x\in\mathbb{C}^n : p(x)=0\: \forall p\in S\right\}$. Certainly, if $V_\mathbb{C}(S)=0$, then $V_\mathbb{R}(S)=0$.

Different sets of polynomials can give rise to the same algebraic variety. For example, if $x$ is a zero of polynomials $f$ and $g$, then it is also a zero of $f+g$ and $fg$. Therefore, $V_\mathbb{C}(S)$ is equivalently given by $V_\mathbb{C}(\langle p_1,\ldots,p_N \rangle)$, where $\langle p_1,\ldots,p_N \rangle$ is the ideal generated by $S=\{p_1,\ldots,p_N\}$,
\begin{equation}
	\langle p_1,\ldots,p_N \rangle\equiv \left\{p\in\mathbb{C}[x_1,\ldots,x_n] : p=\sum_{i=1}^N{r_ip_i},\, r_i\in\mathbb{C}[x_1,\ldots,x_n],\, i=1,\ldots,N \right\}.
\end{equation}
The ideal $\langle p_1,\ldots,p_N\rangle$ is not uniquely generated by $S$. In fact, every ideal $I$ in $\mathbb{C}[x_1,\ldots,x_n]$ is finitely generated, and if $I$ is non-zero, there even exists a unique, distinguished generating set, the reduced Gr\"obner basis (w.r.t. a monomial order, to be defined later), which generates the ideal. 

Gr\"obner bases can be viewed as a generalization of Gaussian elimination for linear systems or the Euclidean algorithm for computing greatest common divisors (see, for example, \cite{adams94}). The notion of a Gr\"obner basis only makes sense after one has defined some monomial order, i.e., a total order on the set of all monic polynomials in a polynomial ring which respects multiplication $(u<v \Rightarrow uw<vw)$, and is a well-ordering (every non-empty set of monomials has a minimal element). For a single variable, the only monomial ordering is $1<x<x^2<x^3\ldots$. For the general multivariate case, one example of a monomial order is the \emph{lexicographic order} (lex). This firstly requires that $x_1>x_2>\ldots>x_n$. For higher degree monomials, the exponents of $x_1$ are compared; in the event of a tie, the exponents of $x_2$ are compared, and so on. Thus, for instance, $x_1^2x_4>x_1x_2^2>x_1x_2x_3^3$. Other examples of monomial orders are the \emph{graded lexicographic order} (grlex) and the \emph{graded reverse lexicographic order} (grevlex). In grlex, the total degree of the monomials are first compared, and ties are broken by applying lex. In grevlex, total degree is first compared; ties are broken by comparing exponents of $x_n$, with smaller exponents regarded as larger in the ordering, followed, if necessary, by comparing exponents of $x_{n-1},x_{n-2}$, etc. Given a monomial order and a polynomial $p\in\mathbb{C}[x_1,\ldots,x_n]$, we denote the largest monomial in $p$ by $\text{lp}(p)$, and its corresponding coefficient by $\text{lc}(p)$. With respect to a given monomial order, a Gr\"obner basis is defined as follows:
\begin{definition}
	A set of non-zero polynomials $G=\{g_1,\ldots,g_t\}$ contained in an ideal $I$, is called a Gr\"obner basis for $I$ iff for all non-zero $f\in I$, there exists $i\in\{1,\ldots,t\}$ such that $\text{\emph{lp}}(g_i)$ divides $\text{\emph{lp}}(f)$.
\end{definition}
A Gr\"obner basis for an ideal is not unique, but the reduced Gr\"obner basis is. This is defined as follows.
\begin{definition}
	A Gr\"obner basis $G=\{g_1,\ldots,g_t\}$ is called a reduced Gr\"obner basis if, for all $i$, $\text{\emph{lc}}(g_i)=1$, and no non-zero monomial in $g_i$ is in the ideal generated by the leading terms of the elements in $G-\{g_i\}$.
\end{definition}

The importance of Gr\"obner bases lies in the following theorem:
\begin{theorem}\label{grobnull}
Let $I$ be an ideal of $\mathbb{R}[x_1,\ldots,x_n]$, and $G$ be its reduced Gr\"obner basis with respect to a monomial order. Then, $V_\mathbb{C}(I)=\emptyset$ if and only if $1\in G$, or equivalently, if and only if $G=\{1\}$.
\end{theorem}
Therefore, the system of real polynomial equations $p_1=0, p_2=0,\ldots,p_N=0$ has no solutions in $\mathbb{C}^n$ iff the reduced Gr\"obner basis for the ideal $\langle p_1,\ldots,p_N \rangle$ is equal to $\{1\}$. In \cite{brierley10}, the authors confirmed that the constellation $\{1,1,1,1\}_2$ does not exist by computing that the polynomial equations defining the constellation have $G=\{1\}$ as the reduced Gr\"obner basis. Thus, we see that we can find out something about the solution set of a system of polynomials (its non-existence for instance), without actually finding the solutions directly. This is the advantage of algebraic geometric methods such as Gr\"obner basis computations. Of course, we should have a way of finding Gr\"obner bases in the first place. Fortunately, there is a very general algorithm, Buchberger's Algorithm \cite{buchberger65}, that can compute the Gr\"obner basis for the ideal $I=\langle p_1,\ldots,p_N \rangle$ in a finite number of steps. Unfortunately, algorithms for computing Gr\"obner bases are not as straightforward as, say, Gaussian elimination. In general, Gr\"obner bases can be very large, and the computational cost of finding one depends very much on the monomial order, the order of the polynomials, and the choice of the so-called $S$-polynomials which appear in the intermediate steps of Buchberger's Algorithm. For instance, it was reported in \cite {brierley10} that the computation of a Gr\"obner basis for the MU constellations $\{5,3,3,3\}_6$ and $\{5,5,4,1\}$ using the package FGb failed because of memory issues (despite having 16GB of memory). Gr\"obner basis computation is still actively researched, and other algorithms such as Faug\`ere's F4 \cite{faugere99} and F5 \cite{faugere02} algorithms are available.

\section{Hilbert's Nullstellensatz, NulLA, and Parrilo's sum-of-squares}
Computation of a Gr\"obner basis may provide an infeasibility certificate for a system of polynomials. However, even that might be more than what we require. There is a beautiful theorem by David Hilbert, which in one particular form states the following:
\begin{theorem}[Hilbert's Nullstellensatz]
	A system of polynomial equations, $p_1(x)=0,\ldots,p_s(x)=0$, has no solutions over an algebraically closed field $\mathbb{K}$ if and only if there exist polynomials $r_1,\ldots,r_s\in\mathbb{K}[x_1,\ldots,x_n]$ such that $1=\sum_{i=1}^s{r_ip_i}$.
\end{theorem}
Therefore, an identity of the form $1=\sum_{i=1}^s{r_ip_i}$ provides an infeasibilty certificate for the system $p_1(x)=0,\ldots,p_s(x)=0$. The maximum degree of the polynomials $r_ip_i$ will be called the degree of the Nullstellensatz certificate. We remark that the Nullstellensatz is intimately linked to Theorem \ref{grobnull} on Gr\"obner bases. We also know that a Nullstellensatz certificate must exist for an infeasible system of polynomials. Therefore, we look to find such a Nullstellensatz refutation for the existence of certain MU constellations. There are in fact systematic ways to search for such refutations, for instance, NulLA (Nullstellensatz Linear Algebra) \cite{deloera08, deloera09}.

The basic idea behind NulLA is quite simple. We fix a tentative degree $d$ for the Nullstellensatz certificate that we wish to find. We then expand the assumed polynomial identity $1=\sum_{i=1}^s{r_ip_i}$ into a linear combination of monomials with degrees less than or equal to $d$. The coefficients of these monomials will be linear expressions in the coefficients defining the unknown polynomials $r_i$. Note that two polynomials over a field are equal if and only if the coefficients of every monomial are equal. Therefore, the identity $1=\sum_{i=1}^s{r_ip_i}$ corresponds to a \emph{linear} system of equations in the coefficients of $r_i$. Solving this linear system then results in two possible outcomes. If the system is consistent, then any solution produces the desired Nullstellensatz certificate of degree $d$. Otherwise, no certificate of degree less than or equal to $d$ exists, and we start afresh with a tentative degree $d+1$. Repeat the process if necessary, until a certificate is found.

The linear systems that appear in NulLA increase very rapidly in size as $d$ increases, and can be huge for reasonably-sized problems, even for $d$ as small as six. We can see this easily. There are $\displaystyle \binom{n+d}{d}$ monomials in $n$ variables of degree $d$ or less. Writing $d_i=d-\text{deg}(p_i)$, there are $\sum_{i=1}^s{\displaystyle \binom{n+d_i}{d_i}}$ unknowns in the linear system to be solved for a Nullstellensatz certificate of degree $d$. Upper bounds on the degree of a Nullstellensatz certificate are known to be doubly-exponential in the number of input polynomials and their degree. However, as pointed out in \cite{deloera08}, fairly low-degree certificates exist for many examples. 

To see how NulLA works out in practice, we look once again at the constellation $\{1,1,1,1\}_2$. The equations defining it are given in \eqref{1111eqn}. After carrying out the NulLA algorithm, we find a Nullstellensatz certificate of degree $6$, given by
\begin{equation}
	\begin{array}{l}
		r_1(x)=\frac{1}{2}\left(-x_1-2x_4^2+x_1x_3^2+x_2x_3x_4-x_2x_3^2x_4-x_2x_4^3\right),\\
	  r_2(x)=\frac{1}{2}\left(-2-x_3+2x_1^2-x_2x_4\right),\\
	  r_3(x)=\frac{1}{2}\left(x_1-x_1x_3^2-x_2x_3x_4\right),\\
	  r_4(x)=\frac{1}{2}x_3,\\
	  r_5(x)=\frac{1}{2}x_2x_4.
	\end{array}
\end{equation}
It is straightforward to verify that $\sum_{i=1}^5{r_ip_i}=1$. We should point out that during the Gr\"obner basis computation carried out in \cite{brierley10}, a slightly different degree 6 Nullstellensatz certificate was produced as a by-product.

Suppose we are looking for a certificate of degree $d$. Then the linear system we need to solve looks like $My=b$. Here, the matrix M has $\displaystyle \binom{n+d}{d}$ rows, one per monomial $x^\alpha$ of degree $d$ or less ($\alpha$ is a multi-index); $M$ also has one column per polynomial of the form $x^\delta f_i$, where $x^\delta$ is a monomial of degree less than or equal to $d-\text{deg}(f_i)$. The vector $b$ has $\displaystyle \binom{n+d}{d}$ entries which are zero everywhere except for the entry corresponding to the constant monomial $x^0$, which is $1$. From this description, one sees that the size the matrix $M$ grows very quickly with the certificate degree $d$.

Still, there are a number of ways to optimize NulLA \cite{deloera08}. Firstly, we note that the system of polynomials $p_1,\ldots,p_s$ defining a MU constellation has only real coefficients. Then it can be shown that it suffices to search for real Nullstellensatz certificates. In other words, a Nullstellensatz certificate $1=\sum_i^s{r_ip_i}$ where $r_i\in\mathbb{C}[x_1\ldots,x_n]$ exists iff there exists a real Nullstellensatz certificate $1=\sum_i^s{r_i'p_i}$ where $r_i'\in\mathbb{R}[x_1,\ldots,x_n]$. Linear algebra can then be carried out over the reals. Secondly, the size of the linear system $My=b$ that has to be solved at each stage of NulLA can be significantly reduced if there are certain symmetries in the system of equations $p_1,\ldots,p_s=0$. For instance, suppose that the set $S=\{p_i\}_{i=1\ldots,s}$ is invariant under the action of a group of permutations $G$ of the variables $x_1,\ldots,x_n$. We denote the image of $p_i$ under $g\in G$ by $g(f_i)$. $G$ also induces an action on the set of monomials of degree $t$. The orbit of a monimial $x^\alpha$ under $G$ is denoted by $O(x^\alpha$), while the orbit of $x^\delta f_i$ is denoted by $O(x^\delta f_i)$. In view of the symmetries captured by $G$, we introduce the matrix equation $\bar{M}\bar{y}=\bar{b}$, where the rows the matrices are indexed by the orbits $O(x^\alpha)$ and the columns indexed by the orbits $O(x^\delta f_i)$. The entries of $\bar{M}$ are defined by
\begin{equation}
	\bar{M}_{O(x^\alpha),O(x^\delta f_i)}=\sum_{x^\gamma f_j\,\in\, O(x^\delta f_i)}{M_{x^\alpha,x^\gamma f_j}}.
\end{equation}
Note that this definition is independent of the choice of $x^\alpha$ in the orbit $O(x^\alpha)$. The vector $\bar{b}$ has zeroes everywhere except for the entry corresponding to $O(1)$. With these definitions, it can be shown that if a solution to $\bar{M}\bar{y}=\bar{b}$ exists, then a solution to $My=b$, exists. 

The above idea may be applicable when dealing with, for instance, the $\{5,5,5,1\}_6$ constellation. In this case, we may choose the first group of vectors to be the computational basis vectors, and the last singleton to be $\frac{1}{\sqrt{6}}(1,1,1,1,1,1)^T$. The remaining two groups of five vectors can be written as
\begin{equation}
	\left\{\frac{1}{\sqrt{6}}\begin{pmatrix}1\\x_{i,j,1}+\im y_{i,j,1}\\x_{i,j,2}+\im y_{i,j,2}\\x_{i,j,3}+\im y_{i,j,3}\\x_{i,j,4}+\im y_{i,j,4}\\x_{i,j,5}+\im y_{i,j,5}\end{pmatrix}\right\}_{i=1,2;j=1,\ldots,5},
\end{equation}
with the index $i$ labelling the group of vectors, and $j$ labelling the five vectors in each group. In this parameterization, there are $100$ real variables, $x_{i,j,k},y_{i,j,k}$, with $k\in\{1,2,3,4,5\}$ labelling the component number. The MU equations are:
\begin{equation}
	\begin{array}{ll}
		x_{i,j,k}^2+y_{i,j,k}^2-1=0&\forall i,j,k,\\
		\left(1+\sum_{k=1}^5{x_{i,j,k}}\right)^2+\left(\sum_{k=1}^5{y_{i,j,k}}\right)^2-6=0&\forall i,j,\\
		\left(1+\sum_{k=1}^5{(x_{i,j,k}x_{i',j',k}+y_{i,j,k}y_{i',j',k})}\right)^2+\left(\sum_{k=1}^5{(x_{i,j,k}y_{i',j',k}-x_{i',j',k}y_{i,j,k})}\right)^2=\left\{\begin{array}{l}6\\0\end{array}\right.&\begin{array}{l}\forall i\neq i',\forall j,j'\\ \forall i=i', \forall j\neq j'\end{array}.
	\end{array}
\end{equation}

It follows that the above set of equations are invariant under simultaneuous permutations {$\{x_{i,j,k};y_{i',j',k'}\}\rightarrow\{x_{g^1(i),g^2(j),g^3(k)};y_{g^1(i'),g^2(j'),g^3(k')}\}$} in each of the three indices, where $g^1,g^2,g^3$ are, respectively, permutations from the permutation groups on $2,5,$ and $5$ objects. In view of this, the linear system $\bar{M}\bar{y}=\bar{b}$ should be significantly simpler than the original system $My=b$. 

A number of other possible ways to improve the efficiency of NulLA are described in \cite{deloera08, deloera09}. These include appending extra polynomial equations from the radical ideal of $\langle p_1,\ldots,p_s\rangle$ to reduce the Nullstellensatz degree, branching the polynomial system into smaller subsystems with the aim of finding lower-degree infeasibility certificates for the smaller subsystems, and using alternative versions of the Nullstellensatz itself. A recent modification to the NulLA algorithm, called FPNulLA (fixed-point NulLA), is also proposed in \cite{deloera09-2}.

So far, we have been trying to show that there are no common complex zeros of the polynomial system $S$ defining a certain MU constellation. It could well be the case that $V_\mathbb{R}(S)=\emptyset$ but $V_\mathbb{C}(S)\neq\emptyset$. In other words, a MU constellation may be non-existent, but no Nullstellensatz certificate can prove it. Unfortunately, the \emph{real} Nullstellensatz is somewhat more complicated than Hilbert's original one. One version says the following:
\begin{theorem}[Real Nullstellensatz]
	Let $p_1,\ldots,p_s\in\mathbb{R}[x_1,\ldots,x_n]$, and let $\Sigma\subset\mathbb{R}[x_1,\ldots,x_n]$ denote the cone of polynomials representable as a sum-of-squares. Then $V_\mathbb{R}(\{p_1,\ldots,p_s\})=\emptyset$ if and only if \,$-1\in\Sigma+\langle p_1,\ldots,p_s \rangle$.
\end{theorem}
The real Nullstellensatz is closely related to Stengle's Positivstellensatz \cite{stengle74}, which deals with the semialgebraic set
\begin{equation}
	S_\text{SA}\equiv\left\{x\in\mathbb{R}^n : p_i(x)=0\,\text{ and }\,q_j(x)\geq 0\;\, \forall i=1,\ldots,s; j=1,\ldots,t\right\},
\end{equation}
where $p_i, q_j\in\mathbb{R}[x_1,\dots,x_n]$. We define $M(\{q_j\})$ to be the set of all finite products of $q_j$, including the empty product, $1$. The cone associated with $\{q_j\}$ is defined as
\begin{equation}
	\text{cone}(q_1,\ldots,q_t)\equiv\left\{g_0+\sum_{k=1}^{r}{g_kb_k} : g_0,\ldots,g_r\in\Sigma, b_1,\ldots,b_r\in M(\{q_j\})\right\}
\end{equation}
The Positivstellensatz reads
\begin{theorem}[Positivstellensatz]
	The semialgebraic set $S_{\text{SA}}=\emptyset$ if and only if $-1\in\text{\emph{cone}}(q_1,\ldots,q_t)+\langle p_1,\ldots,p_s\rangle$.
\end{theorem}
In \cite{parrilo03}, Parrilo described how the search for a Positivstellensatz infeasibility certificate can be rephrased as a hierarchy of semidefinite programs. This essentially involves assuming a certain maximal degree for the infeasibility certificate, and checking the feasbility of the corresponding SDP. If the SDP for a certificate of degree $d$ is not feasible, then one proceeds to check the feasibility of the SDP for a $d+1$ degree certificate.

Direct numerical searches have not been able to find MU constellations such as $\{5,3,3,3\}_6$ and $\{5,5,4,1\}_6$. If such constellations were to exist, it is possible that the corresponding ideal $S=\langle p_i,\ldots,p_s\rangle$ is zero-dimensional, i.e., the set of common zeros, $V_\mathbb{R}(S)$, is finite. In this case, Lasserre \emph{et. al.} \cite{lasserre08-2} have described a numerical algorithm based on semidefinite programming to compute the points on this finite variety. Finally, we remark that infeasibility certificates could be easier to find for systems that are ``more infeasible". From this point of view, one might want to consider directly the case of seven MUBs in $\mathbb{C}^6$ instead of MU constellations.

\section{Conclusion}
A few approaches to the existence problem of four MUBs in $\mathbb{C}^6$ have been described. Current numerical evidence suggests that four MUBs do not exist, but a rigorous \emph{infeasibility certificate} is lacking. One approach to searching for such an infeasibility certificate involves setting up an optimization problem in which the objective function attains the maximum/minimum value in the codomain precisely when a set of four MUBs exists. Then, one can obtain an infeasibility certificate by proving certain \emph{global} bounds on this objective function. The main obstacle in this approach is the absence of convexity, which appears to be crucial; relaxation of the non-convex constraints seems to have the tendency to give trivial global bounds.

A different approach views the problem from the point of view of algebraic geometry. A MU constellation is defined using a set of multivariate polynomial equations, and its non-existence corresponds to the infeasibility of this set of equations. A number of computable infeasibilty certificates are available from the theory of algebraic geometry, but these usually prove the non-existence of complex zeros. Furthermore, the number of variables and constraints required to define a MU constellation is quite large, and poses problems for realistic computation of infeasibilty certificates. One recent idea is to use linear algebra to compute infeasibilty certificates, and it is hoped that with the appropriate refining techniques, the linear systems involved can be solved. Real algebraic geometry is less well-understood, but recent work has linked it to semidefinite programming, which may allow infeasibilty certificates to be computed with a realistic amount of resources.

Special thanks to Philippe Raynal for numerous valuable discussions, and B.-G. Englert for his advice.

\end{document}